# When Coincidence has Meaning: Understanding Emergence Through Networks of Information Token Recurrence

**Markus Luczak-Roesch**[1,2]

[1]Victoria University of Wellington, School of Information Management, New Zealand

[2]Te Pūnaha Matatini, New Zealand's Centre of Research Excellence for Complex Systems

markus.luczak-roesch@vuw.ac.nz

**Abstract:** In this paper I conceptualise a novel approach for capturing coincidences between events that have not necessarily an observed causal relationship. Building on the Transcendental Information Cascades approach I outline a tensor theory of the interaction between rare micro-level events and macro-level system changes. Afterwards, I discuss a number of application areas that are promising candidates for the validation of the theoretical assumptions outlined here in practice. This is preliminary work that is sought to lay the foundation to discover universal mathematical properties of coincidences that have a measurable impact on the macroscopic state of a complex system and are therefore to be considered meaningful.

**Introduction**

Climate change and antimicrobial resistance are two of the largest threats to the safety and wellbeing of people globally [1,2]. Due to the vast number of components and the complexity of the interactions in climate and bacteria, these – amongst many other natural and virtual systems – are considered *complex systems*. Complex systems are prone to unpredictable changes that have no apparent causal explanation in how individual parts of systems change, and that challenge our understanding of why and how complex systems evolve. This unpredictable evolution is known as *emergence* and emergence has so far resisted universal mathematical formulation [3].

Traditionally most efforts to understand evolution in complex systems have focused on causal structures [4,5,6,7,8] or periodic patterns [9,10,11,12,13], or they assume that information is significant if observed frequently [14,15,16]. Such assumptions are at odds with the acausal character of rare, unanticipated coincidences that lead to changes in a system's macroscopic state [3]. Serendipitous discoveries made on citizen science platforms provide an example [17]; it was unpredictable that one user of the Galaxy Zoo project logged in at a specific time, was presented with an image of an uncommon artefact, noticed the artefact, and posted about it, which triggered a change in the behaviour of other users, leading to the confirmation of the novelty of the observation by professional scientists[1]. But how can we rigorously understand what happened? Even if it is possible to retrospectively capture statistical properties of unpredictable changes in a specific domain [18,19,20], there is still no universal model that explains the laws of emergence in general nor of coincidences specifically across particular domains [21,22].

**Foundations and related work**

For those complex systems for which traces of interactions between components (e.g. actors or entities) are available, temporal networks have been proposed as a promising solution to overcome the problem of analytical case- and domain-dependence [22]. The common argument in support of a network approach to complex evolving systems is that network representations allow us to describe qualitatively very different complex phenomena using the same formal model, which is expected to help understand similarities and differences within the macroscopic properties of the different systems more easily, in particular since networks come with well-established mathematical and visual analytical frameworks [41].

Recurrence networks [54,55] and horizontal visibility graphs [66] are two prominent examples of such network approaches to dynamical systems. A recurrence network is the interpretation of a recurrence plot of a time-series with N discrete time steps as an undirected network of N nodes. And horizontal visibility graphs rely on topological features of a single time series for the construction of a temporally ordered network representation of the original data. While being widely adopted to study the dynamics of complex systems by the means of network analysis, the analytical frameworks that come with these and similar state-of-the-art approaches have limited access to the multitude of observable microscopic dimensions that are a characteristic of most complex systems. Or to phrase it differently: existing analytical frameworks capture only one particular information tokenization for one type of data and in their current form do not suit to study systems of systems.

This gap has been articulated by Masuda and Holme in their recent article on the application of temporal networks to capture evolving systems' states [22]. They synthesised: "A limitation […] is the assumption that the entire system can be described by a single system state. […] In many cases, one could argue that it makes more sense to describe different groups in the data as having their own system state." [22] This observation suggests that the aforementioned problem of a lack of comparability of analyses arises not only between different systems but even within single systems. Therefore, any expectation that a single recurrence network or horizontal visibility graph, for example, may be sufficient to fully understand the dynamics and states of a complex time-evolving system has to be challenged. All individual sub-systems may have their very own states reflected in the dynamical properties of different views to the overall system. In other words, when we look at a complex system through different lenses and employ different analytical frameworks to capture its dynamics (looking, for example, both at the continuous time series of a brain signal and the discrete sequence of eye movements), there is currently no analytical framework that allows for integrating the individual insights about each sub-system's evolving states into a macroscopic model of the overall system.

---

[1] See https://en.wikipedia.org/wiki/Hanny's_Voorwerp for details on this case.

In summary, previous research suggests that there is a unique opportunity to understand time-evolving complex systems by combining network science and dynamical systems theory. Yet, no work has resulted in a mathematical model and analytical framework that is universally applicable. With reference to the scope of my work presented here I suggest further that current candidates for such a generic approach do not seem to preserve access to low-level coincidences of individual events underlying macroscopic state changes, which in turn means the approaches will fall short of being able to fully explain emergence.

## A novel data-driven approach to understand emergence

In the following I will outline a novel data-driven approach that brings together temporal networks, tensor decomposition and spectral theory to capture invariant properties of time-evolving complex systems that are inaccessible using existing ergodic theory. The approach shall enable a completely new kind of experiments to derive the unique signatures of meaningful coincidences, those coincidences that have a measurable impact on the macroscopic state of a complex system despite being seemingly random, from observations in data using computational tools and simulations as it is commonly done in experimental mathematics [52,53].

At the core of the proposed approach sits an extension of *Transcendental Information Cascades (TICs)*, a method that transforms any kind of sequential data into a directed temporal network of recurring low-level tokens of information (e.g. words in text, the power spectrum of a signal over a particular time slice or distinct triplets in DNA) [23,24]. TICs have been designed as a kind of analytical middle ground between completely different kinds of sequential data and provide an unprecedented micro- and macro-level view to those data and the temporal dynamics they exhibit in the sense that it is always possible to switch between a macroscopic whole-system-view and a microscopic single-event-view. My hypothesis is: TICs allow us to formalise coincidences and decide computationally which ones are random and which ones carry meaning.

### *Extending Transcendental Information Cascades to capture micro- and macro-level coincidences in sequential data*

To understand the extension of TICs I propose here, it is first necessary to give a brief introduction into TICs. To construct a TIC (cf. Figure 1) all elements of a discrete or discretised (e.g. using time windows over a continuous signal) source data sequence are analysed using a pre-defined information extraction method in chronological order from oldest to newest, and information tokens are extracted that are stored in a distinct set for each sequence element. An information token $i_z$ could be any kind of distinct information that can be identified in the source data by the applied information extraction method $f_k$. Then a network *TC* is created where each sequence element is represented by a node $v_y$ that holds the sequential index and the token set $I_y$ as attributes, and a directed edge $e_x$ is created for any token from those sets to the next node in the sequence that features this particular token in its token set.

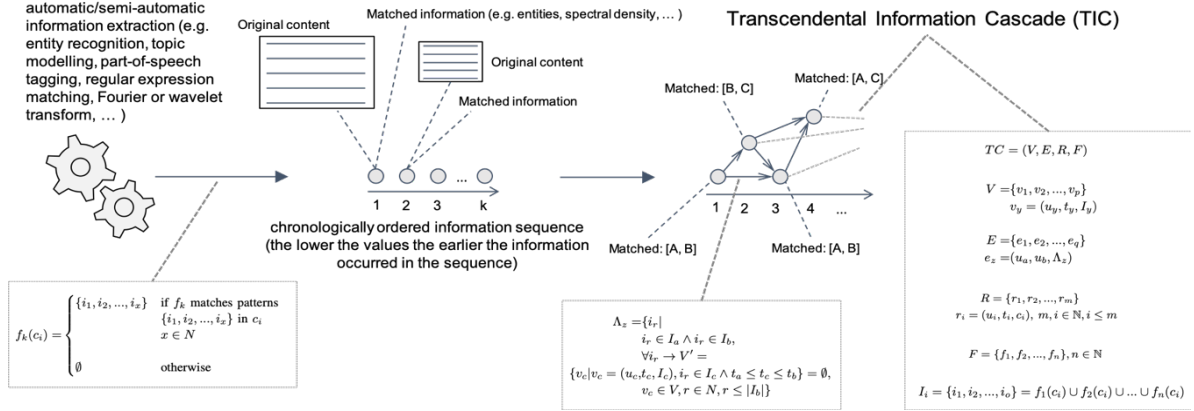

*Figure 1: Overview of the Transcendental Information Cascade (TIC) approach.*

TICs can be considered unique because they allow assessing information theoretic [35,36,37,38,39] and network theoretic properties [40,41] of different kinds of sequential source data (e.g. brain waves, English novels or records of global $CO_2$ measurements) using the same unified temporal network model [24,31]. Key foci of analysis in TICs are recurrences, co-occurrences and bursts. And while state-of-the-art methods would usually construct higher-order complex states from those [14,25,26,27,28,29,30], and thus neglect rare events at lower levels of a system's dynamics (e.g. individual words), TICs by contrast enable exploring both [24,31,32,33,34].

I suggest that it is possible to leverage the characteristic of TICs to capture micro- and macro-level events to derive a universal mathematical definition of coincidences as schematically outlined in Figure 2. From *a* different TICs $TC_i$ ($i \in [1...a]$) that were constructed from the same source data sequence *R* but by applying different information extraction methods $F_i$ (e.g. software source code shared on the stackoverflow question answering site can be tokenised as individual lines of code or abstract syntax trees leading to different tokenisations), one can first derive *a* 2-dimensional vector spaces $Z_i$ that combine the time-series derived from the usual properties (I will refer to these also as features in the remainder of this paper and the symbol $\zeta_m$ denotes a particular feature time-series) that are assessed for TICs at every discrete time step ([24] defined token entropy, Wiener index, token set diversity and token set specificity as the core measures to quantify the information theoretic and network properties of TICs). Additionally, one can understand the *a* adjacency matrices of all TICs as 2-dimensional vector spaces as well. Layering these different groups of 2-dimensional vector spaces results in a number of 3-tensors that all have the time dimension ($t_1, t_2, \ldots t_n$) in common. First, the tensor $\mathfrak{F}$ combining all the feature vector spaces. Second, the tensor $\mathfrak{C}$ combining all the adjacency matrices. And third, a set of tensors $\mathfrak{C}_{\zeta_m}$ that one can derive when using the values of a particular feature time-series $\zeta_m$ as the edge weight in a TIC, which can be represented as the values of an alternative TIC adjacency matrix.

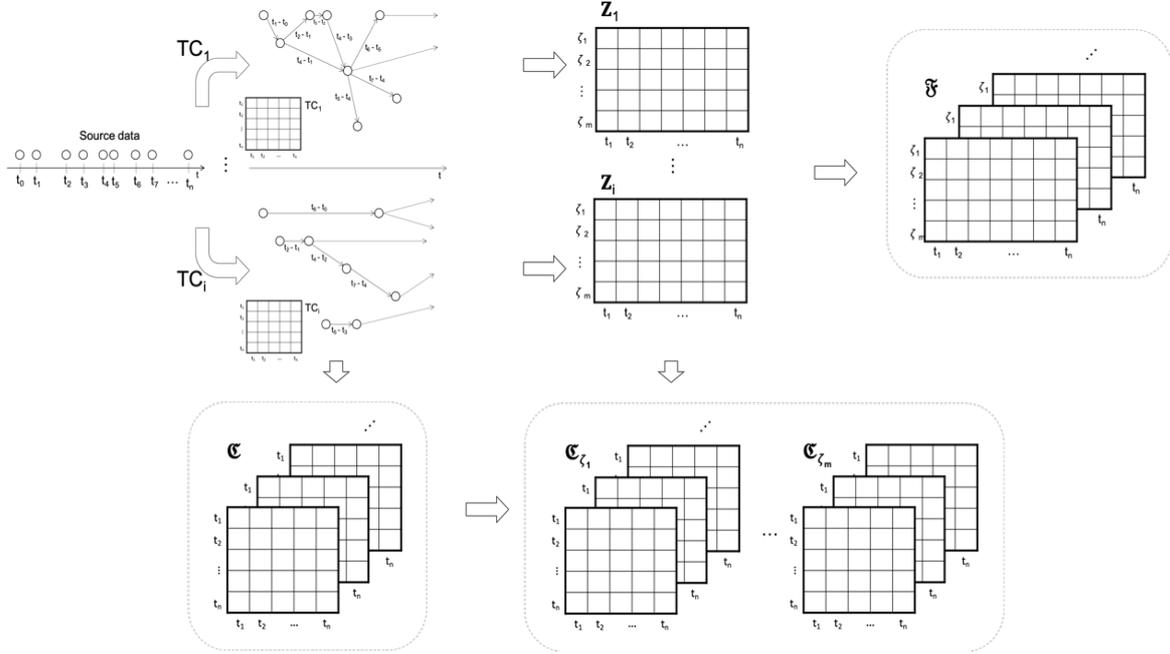

***Figure 2:*** *Conceptual view to how from one source data sequence multiple TICs can be constructed that can be analysed as one integrated macroscopic system through three different kinds of tensor representations.*

Because the focus here is on coincidences, it is necessary to define what constitutes a coincidence in TICs. I define *coincidence* as sets of co-occurring information tokens from any ***$TC_i$*** that are situated in the tail of the respective recurrence and co-occurrence rank-size distributions (i.e. these are tokens that occur and co-occur rarely) over that particular ***$TC_i$*** and that are not part of a causal macroscopic pattern in $\mathfrak{F}$, $\mathfrak{C}$ or $\mathfrak{C}_{\zeta m}$. The latter is necessary to identify latent dependencies between individual TICs that are not visible when looking at one TIC in isolation. A transfer entropy approach that constructs symbol sequences from $\mathfrak{F}$, $\mathfrak{C}$ and $\mathfrak{C}_{\zeta m}$ sliced along the time dimension, and then produces a network of TICs that may be causally related [62] is promising for this task.

By applying canonical polyadic decomposition (CPD) and Tucker decomposition [64,65] to the tensors $\mathfrak{F}$, $\mathfrak{C}$ and $\mathfrak{C}_{\zeta m}$ it is now possible to derive matrices (and in case of Tucker decomposition also a core tensor) that can be considered a characteristic low-dimensional representation of the macroscopic state of the system of all TICs that were constructed for a particular source data sequence (cf. Figure 3). I suggest that the analysis of statistical properties of the spectra of these matrices (e.g. the probability distribution of their eigenvalues) is promising to find out under which conditions a decomposition (or a sub-set of the resulting matrices of a decomposition) may represent a non-random invariant of the dynamics of the underlying TICs.

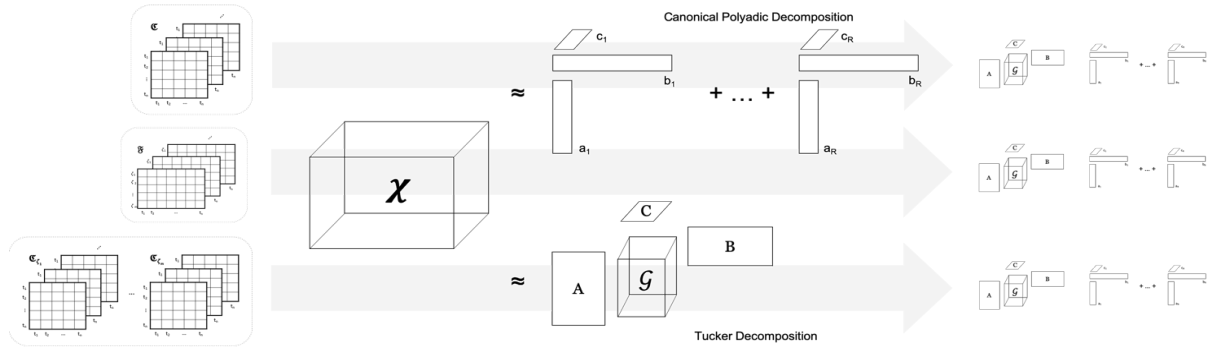

*Figure 3: Conceptual view to how CPD and Tucker decomposition can be applied to the different tensors in order to derive low-dimensional representations of the macroscopic system of all TICs constructed from one source sequence.*

*Defining meaningful coincidences*

It is now necessary to distinguish between meaningless and meaningful coincidences. I define *meaningful coincidences* as those coincidences that have an impact on the macroscopic state of a system of TICs that is observable and non-random [42-44]. Inspired by universality in Wigner matrices [56-60] this works from the assumption that a rare event that that does not change invariant properties of a tensor decomposition can be considered random while a rare event that has an effect on such properties signifies actual meaning. Specifically, I suggest that the statistical properties of the eigenvalues of the matrices of the tensor decompositions suit as the invariant property of interest. Experimental investigation should focus on what stability or variance of the eigenvalue statistics in one or more tensor dimensions mean with reference to an underlying rare event in the source data, and how this can be formally expressed so it is possible to recover the root cause of a macroscopic state change.

*Applying the new approach to real and simulated data*

To validate the operationalisation of meaningful coincidences it is required to run experiments with TICs constructed from various real-world datasets and simulated datasets. The idea is to selectively drop those tokens from the TIC construction process that are found to be coincidental as per aforementioned definition, and then re-run the tensor construction, tensor decomposition and spectral analysis to see what difference it makes for the macroscopic system state if a particular token is considered as if it hasn't been observed. If dropping a token does not change the spectral properties of the dimension matrices it can be considered random noise while it otherwise is considered to have meaningful impact.

Through initial experiments I identified four domains in which an understanding of coincidences promises new insights about emergence and obtained relevant data sets. In particular I started studying the Berkeley Earth Surface Temperatures [47] as an example for climate data; the BItsliced Genomic Signature Index for genomic data of bacteria [48], data about software engineering practices from the SOTorrent dataset [49], and language use data from the Bequia Corpus [50,51]. As a next step it is necessary to use the method described here to access coincidences in these data that have not been investigated so far because they are inaccessible to state-of-the-art methods. This will then allow to assess the meaning of these coincidences and address questions like: Which climate indicators feature coinciding event markers and does that provide new insights into climate change? Which coincidences at the genomic level of bacteria may relate to antimicrobial resistance? Do software engineering practices coincide in different programming languages leading to wide adoption of new programming paradigms or patterns? Do coincidences in socio-linguistic features help explain language diversity and evolution?

Further application areas for which data is reasonably accessible and for which the proposed analytical approach promises valuable new insights are:

- Neuroscience data (e.g. EEG brain wave recordings): analysis could focus on detecting seizures and/or understanding sleep patterns.
- Literature (in English and other languages): analysis could focus on studying the evolution of different languages at the level of their symbolic representations (written form) blended with the semantics.
- Autobiographical memories of individuals (spoken or written life stories): analysis could focus on understanding personality shifts and/or indicators of dementia.
- Social media data: analysis could focus on the spread of misinformation and/or the emergence of trending topics.
- Data from peer-production systems (e.g. Wikipedia editing activity and discussions): analysis could focus on dynamics in collective memory and/or biases in crowdsourced information.

For all application areas it will be necessary to compare the results from applying TICs to the different datasets with the results alternative methods for the study of dynamical systems produce for the same data. The focus of study should be on alternative theories that allow quantifying whether a system at hand can be considered chaotic and whether it features attractors and/or periodicity [42,43,45,46,54,63]. Specifically, I suggest assessing the Renyi dimensions and the Lyapunov exponents, perform bifurcation and recurrence quantification, and determine the permutation entropy of the time-series that are part of the different vector spaces constructed from the information theoretic and network features of TICs. To understand relationships between the different features I suggest the transfer entropy of combinations of time-series [14,61].

The final question is whether a chosen tokenization actually leads to a non-random observation or whether what is captured as a meaningful coincidence can also be found in random data. I propose simulations with TICs constructed from Brownian motion and white noise, because these models lead to data that is not meaningfully complex in terms of the characteristic distributions, and from the Erdős–Rényi model and the Lorentz system, because these simulate systems with well-understood and meaningful complexity.

**Conclusion**

This brief conceptual paper lays out the foundations of a new attempt to mathematically formalise laws of emergence. The major goal of this line of research is to develop an understanding of emergence in complex systems mathematically and empirically, focussing specifically on coincidences – seemingly random co-occurrences of independent events on different levels of abstractions of any considered quantity (i.e. varying quantities that have different units/ways of being measured).

I conceptualised a method that shall be capable of (1) tracking patterns across analytical levels and longitudinal scales of emergent phenomena – those that are locally dependent (e.g. the shift of language diversity over the course of an hour long conversation) and those that are locally independent but interdependent over very long time periods (accumulation of individually tolerable mutations that lead to shifts in a phenotype over long time periods) – and (2) making the analysis comparable even if the data are of different types and from different domains.

The theoretical contribution of this work builds upon and advances work in multilayer networks [67-69]. Existing approaches for the study of the structure and dynamics of multilayer networks heavily leverage spectral properties, but either focus on comparing network structures of different layers (e.g. to see if a vertex with high centrality in one layer is less central in another) or exploit explicit links between layers (e.g. links resulting from causal structures in the observed system) to characterize cross-layer multiplexity. In contrast to this, the Transcendental Information Cascades approach focuses on temporal coincidence as a kind of hidden cross-layer multiplexity that occurs spontaneously and for which we lack analytical methods.

As mentioned before, an approach like the one described here is highly desired [41] but has not been developed so far [22,70]. This work has the potential to fill this gap and be a major step towards universal laws of emergence, open new opportunities for investigations that link network science with dynamical systems theory, and provide new insights about chaos and randomness.

**References**

[1] Antimicrobial resistance – an imminent threat to Aotearoa, New Zealand. From the Office of the PMCSA, endorsed by the Science Advisory Network. November 2018. Retrieved from https://www.pmcsa.ac.nz/2018/11/02/antimicrobial-resistance/ on 04/02/2019.

[2] New Zealand's changing climate and oceans: The impact of human activity and implications for the future. An assessment of the current state of scientific knowledge by the Office of the Chief Science Advisor. July 2013. Retrieved from https://www.pmcsa.org.nz/wp-content/uploads/New-Zealands-Changing-Climate-and-Oceans-report.pdf on 04/02/2019.

[3]     Goldstein, J., 2011. Emergence in complex systems. The Sage Handbook of Complexity and Management, ISBN 9781847875693, pp.65-78.

[4]     Gruhl, D., Guha, R., Liben-Nowell, D. and Tomkins, A., 2004, May. Information diffusion through blogspace. Proceedings of the 13th international conference on World Wide Web.

[5]     Danescu-Niculescu-Mizil, C., West, R., Jurafsky, D., Leskovec, J. and Potts, C., 2013, May. No country for old members: User lifecycle and linguistic change in online communities. Proceedings of the 22nd international conference on World Wide Web.

[6]     Watts, D.J., 1999. Networks, dynamics, and the small-world phenomenon. American Journal of sociology, 105(2), pp.493-527.

[7]     Andres, J., 2010. On a conjecture about the fractal structure of language. Journal of Quantitative Linguistics, 17(2), pp.101-122.

[8]     Milička, J., 2014. Menzerath's Law: The Whole is Greater than the Sum of its Parts. Journal of Quantitative Linguistics, 21(2), pp.85-99.

[9]     Zhang, J. and Small, M., 2006. Complex network from pseudoperiodic time series: Topology versus dynamics. Physical review letters, 96(23), p.238701.

[10]    Small, M., 2005. Applied nonlinear time series analysis: applications in physics, physiology and finance (Vol. 52). World Scientific.

[11]    Donges, J.F., Zou, Y., Marwan, N. and Kurths, J., 2009. Complex networks in climate dynamics. The European Physical Journal-Special Topics, 174(1), pp.157-179.

[12]    Donges, J.F., Zou, Y., Marwan, N. and Kurths, J., 2009. The backbone of the climate network. EPL (Europhysics Letters), 87(4), p.48007.

[13]    Yang, J., McAuley, J., Leskovec, J., LePendu, P. and Shah, N., 2014, April. Finding progression stages in time-evolving event sequences. In Proceedings of the 23rd international conference on World wide web (pp. 783-794). ACM.

[14]    Borge-Holthoefer, J., Perra, N., Gonçalves, B., González-Bailón, S., Arenas, A., Moreno, Y. and Vespignani, A., 2016. The dynamics of information-driven coordination phenomena: A transfer entropy analysis. Science advances, 2(4), p.e1501158.

[15]    Mei, Q. and Zhai, C., 2005, August. Discovering evolutionary theme patterns from text: an exploration of temporal text mining. In Proceedings of the eleventh ACM SIGKDD international conference on Knowledge discovery in data mining (pp. 198-207). ACM.

[16]    Subašić, I. and Berendt, B., 2013. Story graphs: Tracking document set evolution using dynamic graphs. Intelligent Data Analysis, 17(1), pp.125-147.

[17]    Tinati, R., Van Kleek, M., Simperl, E., Luczak-Rösch, M., Simpson, R. and Shadbolt, N., 2015, April. Designing for citizen data analysis: a cross-sectional case study of a multi-domain citizen science platform. In Proceedings of the 33rd Annual ACM Conference on Human Factors in Computing Systems (pp. 4069-4078).

[18] Danishvar, M., Mousavi, A. and Broomhead, P., 2018. EventiC: A Real-Time Unbiased Event-Based Learning Technique for Complex Systems. IEEE Transactions on Systems, Man, and Cybernetics: Systems.

[19] Hamilton, W.L., Leskovec, J. and Jurafsky, D., 2016. Diachronic word embeddings reveal statistical laws of semantic change. arXiv preprint arXiv:1605.09096.

[20] Vajna, S., Tóth, B. and Kertész, J., 2013. Modelling bursty time series. New Journal of Physics, 15(10), p.103023.

[21] Diaconis, P., & Mosteller, F. (1989). Methods for Studying Coincidences. Journal of the American Statistical Association, 84(408), 853–861. https://doi.org/10.2307/2290058

[22] Masuda, N. and Holme, P., 2019. Detecting sequences of system states in temporal networks. Scientific reports, 9(1), p.795.

[23] Luczak-Roesch, M., Tinati, R. and Shadbolt, N., 2015, May. When resources collide: Towards a theory of coincidence in information spaces. In Proceedings of the 24th International Conference on World Wide Web. ACM.

[24] Luczak-Roesch, M., O'Hara, K., Tinati, R. and Dinneen, J. D. 2018. What an entangled Web we weave: An information-centric approach to socio-technical systems. Minds & Machines, 28: 709, https://doi.org/10.1007/s11023-018-9478-1.

[25] Kleinberg, J., 2003. Bursty and hierarchical structure in streams. Data Mining and Knowledge Discovery, 7(4), pp.373-397.

[26] Barabasi, A.L., 2005. The origin of bursts and heavy tails in human dynamics. Nature, 435(7039), pp.207-211.

[27] Eckmann, J.P., Kamphorst, S.O. and Ruelle, D., 1987. Recurrence plots of dynamical systems. EPL (Europhysics Letters), 4(9), p.973.

[28] Donner, R.V., Small, M., Donges, J.F., Marwan, N., Zou, Y., Xiang, R. and Kurths, J., 2011. Recurrence-based time series analysis by means of complex network methods. International Journal of Bifurcation and Chaos, 21(04), pp.1019-1046.

[29] Donges, J. F., Schleussner, C.-F., Siegmund, J. F., & Donner, R. V. (2016). Event coincidence analysis for quantifying statistical interrelationships between event time series On the role of flood events as triggers of epidemic outbreaks. Eur. Phys. J. Special Topics, 225, 471–487. https://doi.org/10.1140/epjst/e2015-50233-y

[30] Griffiths, T. L., & Tenenbaum, J. B. (2007). From mere coincidences to meaningful discoveries. Cognition, 103(2), 180–226. https://doi.org/10.1016/j.cognition.2006.03.004

[31] Luczak-Roesch, M., Tinati, R., Van Kleek, M. and Shadbolt, N., 2015, August. From coincidence to purposeful flow? Properties of Transcendental Information Cascades. In Proceedings of the 2015 IEEE/ACM International Conference on Advances in Social Networks Analysis and Mining 2015. ACM.

[32] Tinati, R., Luczak-Roesch, M. and Hall, W., 2016, April. Finding Structure in Wikipedia Edit Activity: An Information Cascade Approach. In Proceedings of the 25th International Conference Companion on World Wide Web (pp. 1007-1012). International World Wide Web Conferences Steering Committee.

[33] Luczak-Roesch, M., Grener, A. and Fenton, E., 2018. Twenty Thousand Leagues Above the Book: An Interactive Visual Analytics Approach to Literature. In Proceedings of the International Conference on Supporting Group Work (GROUP), ACM. DOI: 10.1145/3148330.3154507.

[34] Luczak-Roesch, M., Grener, A. and Fenton, E., 2018. Not-So-Distant Reading: A Dynamic Network Approach to Literature. it-Information Technology, 60(1), pp.29-40.

[35] Rabiner, L.R., 1989. A tutorial on hidden Markov models and selected applications in speech recognition. Proceedings of the IEEE, 77(2), pp.257-286.

[36] Anick, D., Mitra, D. and Sondhi, M.M., 1982. Stochastic Theory of a Data-Handling System with Multiple Sources. Bell Labs Technical Journal, 61(8), pp.1871-1894.

[37] Schirdewan, A., Gapelyuk, A., Fischer, R., Koch, L., Schütt, H., Zacharzowsky, U., Dietz, R., Thierfelder, L. and Wessel, N., 2007. Cardiac magnetic field map topology quantified by Kullback-Leibler entropy identifies patients with hypertrophic cardiomyopathy. Chaos: An Interdisciplinary Journal of Nonlinear Science, 17(1), p.015118.

[38] Shannon, C.E., 1949. Communication theory of secrecy systems. Bell Labs Technical Journal, 28(4), pp.656-715.

[39] Kullback, S., 1997. Information theory and statistics. Courier Corporation.

[40] Williams, M.J. and Musolesi, M., 2016. Spatio-temporal networks: reachability, centrality and robustness. Open Science, 3(6), p.160196.

[41] Holme, P. and Saramäki, J., 2012. Temporal networks. Physics reports, 519(3), pp.97-125.

[42] Schnorr, C.P., 1972, May. The process complexity and effective random tests. In Proceedings of the fourth annual ACM symposium on Theory of computing (pp. 168-176).

[43] Martin-Löf, P., 1966. The definition of random sequences. Information and control, 9(6).

[44] Gács, P., 2005. Uniform test of algorithmic randomness over a general space. Theoretical Computer Science, 341(1-3), pp.91-137.

[45] Webber Jr, C.L. and Zbilut, J.P., 2005. Recurrence quantification analysis of nonlinear dynamical systems. Tutorials in contemporary nonlinear methods for the behavioral sciences, pp.26-94.

[46] Boeing, G. 2016. "Visual Analysis of Nonlinear Dynamical Systems: Chaos, Fractals, Self-Similarity and the Limits of Prediction." Systems, 4 (4), 37. doi:10.3390/systems4040037

[47] Rohde, R., Muller, R., Jacobsen, R., Perlmutter, S., Rosenfeld, A., Wurtele, J., Curry, J., Wickhams, C. and Mosher, S., 2013. Berkeley Earth Temperature Averaging Process. Geoinfor Geostat: An Overview 1: 2. of, 13, pp.20-100.

[48] Bradley, P., den Bakker, H.C., Rocha, E.P., McVean, G. and Iqbal, Z., 2019. Ultrafast search of all deposited bacterial and viral genomic data. Nature Biotechnology, 37(2).

[49] Baltes, S., Treude, C. and Diehl, S., 2018. SOTorrent: Studying the Origin, Evolution, and Usage of Stack Overflow Code Snippets. arXiv preprint arXiv:1809.02814.

[50] Walker, J.A. and Meyerhoff, M., 2006. Zero copula in the eastern Caribbean: Evidence from Bequia. American Speech, 81(2), pp.146-163.

[51] Walker, J.A. and Meyerhoff, M., 2015. Bequia English. Further Studies in the Lesser-Known Varieties of English, p.128.

[52] Campbell, D., Farmer, D., Crutchfield, J. and Jen, E., 1985. Experimental mathematics: the role of computation in nonlinear science. Communications of the ACM, 28(4), pp.374-384.

[53] Bailey, D.H. and Borwein, J.M., 2005. Experimental mathematics: Examples, methods and implications. Notices of the AMS, 52(5), pp.502-514.

[54] Bradley, E. and Kantz, H., 2015. Nonlinear time-series analysis revisited. Chaos: An Interdisciplinary Journal of Nonlinear Science, 25(9), p.097610.

[55] Donner, R.V., Small, M., Donges, J.F., Marwan, N., Zou, Y., Xiang, R. and Kurths, J., 2011. Recurrence-based time series analysis by means of complex network methods. International Journal of Bifurcation and Chaos, 21(04), pp.1019-1046.

[56] Wigner, E.P., 1958. On the distribution of the roots of certain symmetric matrices. Ann. Math, 67(2), pp.325-327.

[57] Mello, P.A. and Baranger, H.U., 1995. Electronic transport through ballistic chaotic cavities: an information theoretic approach. Physica A: Statistical Mechanics and its Applications, 220(1-2), pp.15-23.

[58] Krbálek, M. and Seba, P., 2000. The statistical properties of the city transport in Cuernavaca (Mexico) and random matrix ensembles. Journal of Physics A: Mathematical and General, 33(26), p.L229.

[59] Bourgade, P., Yau, H.T. and Yin, J., 2018. Random band matrices in the delocalized phase, I: Quantum unique ergodicity and universality. arXiv preprint arXiv:1807.01559.

[60] Ajanki, O.H., Erdős, L. and Krüger, T., 2017. Universality for general Wigner-type matrices. Probability Theory and Related Fields, 169(3-4), pp.667-727.

[61] Schreiber, T., 2000. Measuring information transfer. Physical review letters, 85(2), p.461.

[62] Staniek, M. and Lehnertz, K., 2008. Symbolic transfer entropy. Physical Review Letters, 100(15), p.158101.

[63] Bandt, C. and Pompe, B., 2002. Permutation entropy: a natural complexity measure for time series. Physical review letters, 88(17), p.174102.

[64] Kolda, T.G. and Bader, B.W., 2009. Tensor decompositions and applications. SIAM review, 51(3), pp.455-500.

[65] Rabanser, S., Shchur, O. and Günnemann, S., 2017. Introduction to tensor decompositions and their applications in machine learning. arXiv preprint arXiv:1711.10781.

[66] Luque, B., Lacasa, L., Ballesteros, F. and Luque, J., 2009. Horizontal visibility graphs: Exact results for random time series. Physical Review E, 80(4), p.046103.

[67] Kivelä, M., Arenas, A., Barthelemy, M., Gleeson, J.P., Moreno, Y. and Porter, M.A., 2014. Multilayer networks. Journal of complex networks, 2(3), pp.203-271.

[68] Boccaletti, S., Bianconi, G., Criado, R., Del Genio, C.I., Gómez-Gardenes, J., Romance, M., Sendina-Nadal, I., Wang, Z. and Zanin, M., 2014. The structure and dynamics of multilayer networks. Physics Reports, 544(1), pp.1-122.

[69] De Domenico, M., Solé-Ribalta, A., Cozzo, E., Kivelä, M., Moreno, Y., Porter, M.A., Gómez, S. and Arenas, A., 2013. Mathematical formulation of multilayer networks. Physical Review X, 3(4), p.041022.

[70] Mason A. Porter. (2019). Nonlinearity + Networks: A 2020 Vision. arxiv preprint 1911.03805, https://arxiv.org/abs/1911.03805.